\documentclass[conference, final]{IEEEtran}
\IEEEoverridecommandlockouts                              
\usepackage{cite}
\usepackage{amsmath,amssymb, amsfonts, amsthm}
\usepackage{algpseudocode} 
\usepackage{algorithm} 
\usepackage{algorithmicx} 
\usepackage{graphicx}
\graphicspath{{fig/}{../fig/}}
\usepackage{cuted}
\usepackage{multicol}
\usepackage[latin1]{inputenc}
\usepackage{lipsum}  
\usepackage{textcomp}
\def\BibTeX{{\rm B\kern-.05em{\sc i\kern-.025em b}\kern-.08em
    T\kern-.1667em\lower.7ex\hbox{E}\kern-.125emX}}
    
\usepackage{color}

\providecommand{\com[2]}{\begin{tt}[#1: #2]\end{tt}}
\usepackage[normalem]{ulem}

\newcommand{\beq}{\begin{equation}}
\newcommand{\eeq}{\end{equation}}
\newcommand{\beqa}{\begin{eqnarray}}
\newcommand{\eeqa}{\end{eqnarray}}
\newcommand{\beqan}{\begin{eqnarray*}}
\newcommand{\eeqan}{\end{eqnarray*}}
\newcommand{\bef}{\begin{figure}}
\newcommand{\enf}{\end{figure}}

\definecolor{magenta}{cmyk}{0.1, 1, 0, 0}
\definecolor{greenedu}{cmyk}{1, 0, 1, 0.1}
\definecolor{cyanedu}{cmyk}{1, 0, 0, 0.1}
\definecolor{brown}{cmyk}{0.2, 0.63, 0.84, 0.2}
\definecolor{BZcolor}{cmyk}{0, 0.35, 1, 0}

\newcommand{\bi}{\begin{itemize}}
\newcommand{\ei}{\end{itemize}}
\newcommand{\bc}{\begin{center}}
\newcommand{\ec}{\end{center}}
\newcommand{\ba}{\begin{array}}
\newcommand{\ea}{\end{array}}
\newcommand{\be}{\begin{equation}}
\newcommand{\ee}{\end{equation}}
\newcommand{\beno}{\begin{equation*}}
\newcommand{\eeno}{\end{equation*}}
\newcommand{\beqna}{\begin{eqnarray}}
\newcommand{\eeqna}{\end{eqnarray}}
\newcommand{\bd}{\begin{displaymath}}
\newcommand{\ed}{\end{displaymath}}
\newcommand{\beqnd}{\begin{eqnarray*}}
\newcommand{\eeqnd}{\end{eqnarray*}}

\renewcommand{\ni}{\noindent}

\newcommand{\emp}[1]{$\text{EMP}_#1$}

\newtheorem{theorem}{\bf Theorem}[section]

\newtheorem{proposition}{\bf Proposition}[section]

\newtheorem{definition}{\bf Definition}

\definecolor{red}{rgb}{1,0,0}

\definecolor{blu}{rgb}{0,0,1}

\definecolor{gre}{rgb}{0,0.7,0.3}

\definecolor{bz}{rgb}{1,.7,0}

\renewcommand\qedsymbol{\hfill$\blacksquare$}
\makeatletter
\renewenvironment{proof}[1][\proofname]{\par
  \normalfont \topsep6\p@\@plus6\p@\relax
  \trivlist
  \item[\hskip\labelsep
        \itshape
    #1\@addpunct{.}]\ignorespaces
}{%
  \nolinebreak\qedsymbol\endtrivlist\@endpefalse
}
\makeatother

\begin{document}
                  
\title{Extending identifiability results from isolated networks to embedded networks}

\author{
\IEEEauthorblockN{Eduardo Mapurunga}
\IEEEauthorblockA{Member, IEEE.
}\and
\IEEEauthorblockN{Michel Gevers}
\IEEEauthorblockA{Life Fellow, IEEE.
}\and
\IEEEauthorblockN{Alexandre S. Bazanella}
\IEEEauthorblockA{Senior Member, IEEE.
}
\thanks{Eduardo Mapurunga and Alexandre S.  Bazanella are with the Data Driven Control Group, Department of Automation and Energy, 
Universidade Federal do Rio Grande do Sul, Porto Alegre-RS, Brazil, \{eduardo.mapurunga, bazanella\}@ufrgs.br}
\thanks{Michel Gevers is with the Institute of Information and Communication Technologies, Electronics and Applied Mathematics (ICTEAM), UCLouvain, Louvain la Neuve, Belgium,
michel.gevers@uclouvain.be.}
\thanks{This work  was supported in part by Conselho Nacional de Desenvolvimento Cient{\' i}fico e Tecnol{\' o}gico (CNPq),
and by Wallonie-Bruxelles International (WBI).}
}

\maketitle

\begin{abstract}
 This paper deals with the design of Excitation and Measurement Patterns (EMPs) for the identification of dynamical networks, when the objective is to identify only a subnetwork embedded in a larger network. Recent results have shown how to construct EMPs that guarantee  identifiability for a range of networks with specific graph topologies, such as trees, loops, or Directed Acyclic Graphs (DAGs). However, an EMP that is valid for the identification of a subnetwork taken in isolation may no longer be valid when that subnetwork is embedded in a larger network. Our main contribution is to exhibit conditions under which it does remain valid, and to propose ways to enhance such EMP when these conditions are not satisfied.
\end{abstract}



\section{Introduction}

This paper deals with the design of Excitation and Measurement Patterns (EMP) for the identification of dynamical networks. 
The network framework used here was introduced in \cite{vandenhof-dankers-heuberger-etal-identification-2013}, where signals are represented as nodes of the network which are related to other nodes through transfer functions.  
These networks can be interpreted as directed graphs (or digraphs) where the transfer functions, also called modules, are the edges of the graph and the node signals are the vertices.

In \cite{vandenhof-dankers-heuberger-etal-identification-2013} and in subsequent contributions, it was assumed that either all nodes are excited or all nodes are measured. 
A breakthrough was made in \cite{bazanella-gevers-hendrickx-network-2019}, where the first identifiability results were obtained for networks where it is not assumed that all nodes are either excited or measured. This is the situation considered in the present paper.  Throughout this paper we refer to such networks as ``networks with partial excitation and measurement''.

Since the publication of \cite{bazanella-gevers-hendrickx-network-2019}, the search has been for the construction of Excitation and Measurement Patterns (EMP) which guarantee network identifiability and that are preferably sparse. 
An EMP defines which nodes are excited and which nodes are measured.  
It is called valid for a given network when it guarantees generic identifiability of that network. 
It is sparse when it is valid and the sum of the number of excited nodes and measured nodes, called cardinality of the EMP, is kept small.
It is called minimal when it  is valid with minimal cardinality. Precise definitions will be given in Section~\ref{defnot}. 

In this paper, we focus on the synthesis of EMPs for networks with partial excitation and measurement. We briefly review the existing results so far. They are all based on the graph-theoretic framework developed in \cite{hendrickx-gevers-bazanella-identifiability-2019}. 
In \cite{bazanella-gevers-hendrickx-network-2019}, the first results were obtained for the generic  identification of networks with specific graph topologies, namely trees and loops.
Subsequent results for the construction of valid EMPs have been obtained 
				for some classes of parallel networks in \cite{mapurunga-identifiability-2021}, 
for DAGs in \cite{mapurunga-gevers-baza-CDC2022}, for isolated loops in  \cite{mapurunga-gevers-bazanella-necessary-2022}, where a necessary and sufficient condition was derived leading to a minimal EMP. 
The construction of a valid EMP  for the identification of a single module was proposed  in 
				\cite{shi-cheng-vandenhof-single-2021}. The paper \cite{cheng-shi-vandenhof-necessary-2023}, while not proposing a synthesis method, presented a new set of necessary conditions for network identifiability in the context of partial excitation and measurement. For this same context, a different approach was proposed in \cite{legat-hendrickx-combinatorial-characterization-2023, legat-hendrickx-pathbased-2021a}; it is not based on the  synthesis of a valid EMP, but on an efficient and fast algorithm that allows to check the validity of large numbers of EMPs.


A first new result of the present paper is a necessary and sufficient condition for the generic identifiability of another class of networks with a specific topology, namely Parallel Paths Networks (PPN), leading to the construction of minimal  EMPs for such structures. 
However, constructing a valid EMP for a new class of networks with specific structure is not the  main object of the present  paper. Instead, we consider a novel approach to the identification of networks, by adressing the following problem. 

It is often the case that one wants to identify a subnetwork that is part of a larger network. 
All the results summarized above lead to the construction of an EMP that is valid for the subnetwork under consideration (whether its graph is a tree, a loop, a PPN, or any other structure) when it is treated in isolation. However, an EMP that is valid for such subnetwork considered in isolation may no longer be valid when  it is embedded in a larger network, as we illustrate 
 in Section~\ref{isolemb}. 

Our main contribution will be to present a set of sufficient conditions under which an EMP that is valid for a subnetwork in isolation remains valid when that subnetwork is embedded in a larger network. Our theorem will also indicate how one can often complement an EMP that is valid for an isolated subnetwork so that the augmented EMP remains valid for  that subnetwork when it is embedded into the larger network. 

The paper is organized as follows. In Section~\ref{defnot} we introduce the notations and the main concepts about generic identifiability of networks used in this paper, and we recall the necessary conditions for the generic identifiability of any network. In Section~\ref{specific} we first recall existing necessary and sufficient conditions for the identifiability of trees and loops, leading to minimal EMPs for these structures. We then present necessary and sufficient conditions for the generic identifiability of Parallel Paths Networks. In Section~\ref{isolemb} we illustrate why an EMP that is valid for a subnetwork treated in isolation may no longer be valid when that subnetwork is embedded in a larger network. Our main result is in Section~\ref{mainresult}: we present two sets of conditions under which an EMP that is valid for a subnetwork treated in isolation remains valid when that subnetwork is embedded. We illustrate this result with an example, which shows how the main theorem can be used to construct a valid EMP for the whole network by combining EMPs that are valid, on the basis of the main theorem,  for subnetworks that, together, compose the whole network.
 
\section{Definitions, Notations and Preliminaries}\label{defnot}

\label{sec:defs}

In this section, we introduce the dynamic networks with partial excitation and measurement that we  deal with in this paper. 
We recall the necessary conditions for generic identifiability that were derived in \cite{bazanella-gevers-hendrickx-network-2019}, and  we also recall the concept of  a valid Excitation and Measurement Pattern (EMP).

We consider dynamic networks composed of $n$ nodes (or vertices) which represent internal scalar signals $\left\lbrace w_k(t) \right\rbrace$ for $k \in  \{1, 2, \dots, n\}$.
These nodes are interconnected by discrete time transfer functions, represented by edges, which are entries of a \emph{network matrix} $G(z)$.
The dynamics of the network is given by the following equations:
\begin{subequations}
	\begin{align}
	w(t) &= G(z)w(t) + Br(t), \label{eq:dynet1} \\
	y(t) &= Cw(t), \label{eq:dynet2}
	\end{align}
\end{subequations}
where $w(t) \in \mathbb{R}^n$ is the node vector, $r(t) \in \mathbb{R}^m$ is the input vector, and $y(t) \in \mathbb{R}^p$ is the set of measured nodes, considered as the output vector of the network.
The matrix $B \in \mathbb{Z}_2^{n \times m}$, where $\mathbb{Z}_2 \triangleq \{0, 1\}$, 
is a binary selection matrix with a single $1$ and $n-1$ zeros  in each column; it selects which  nodes are excited.
Similarly, $C \in \mathbb{Z}_2^{p \times n}$ is a matrix with a single $1$ and $n-1$ zeros in each row that selects which nodes are measured. 

 To each network matrix $G(z)$ we associate a directed graph (called digraph) $\mathcal{G}$ defined by the tuple $(\mathcal{V}, \mathcal{E})$, where $\mathcal{V}$ is the set of vertices and $\mathcal{E} \subseteq \mathcal{V} \times \mathcal{V}$ is the set of edges. The digraph $\mathcal{G}$ defines the topology of the network.
	The edges are associated to transfer functions, which define the relationships between the nodes.
	We adopt the terminology $j \to i$ to denote an edge $(j, i)$.
	Node $j$ is called an in-neighbor of node $i$, and node $i$ is an
	out-neighbor of node $j$. There is an edge $(j, i)$ associated with the transfer function $G_{ij}(z)$
	of the network matrix only if $G_{ij}(z)$ is nonzero. 
				In this paper, we consider only \emph{connected} graphs, meaning that every node can be reached by at least one other node in the network.
				A node that has no in-neighbors is called a \emph{source}, while a node that has
				no out-neighbors is called a \emph{sink}. 

For the digraph $\mathcal{G}$ associated to the network matrix $G(z)$ we introduce the following notations.
\begin{itemize}
	\item $\mathcal{V}$ -- the set of all $n$ nodes;
	\item $\mathcal{B}$ -- the set of excited nodes, defined by $B$ in (\ref{eq:dynet1});
	\item $\mathcal{C}$ -- the set of measured nodes, defined by $C$ in (\ref{eq:dynet2});
	\item $\mathcal{F}$ -- the set of sources;
	\item $\mathcal{D}_F$ -- the set of dources: see Definition \ref{def:dource} below;
	\item $\mathcal{S}$ -- the set of sinks;	
	\item $\mathcal{D}_S$ -- the set of dinks: see Definition \ref{def:dource} below;
\end{itemize}
The dources and dinks are defined as follows (see \cite{mapurunga-gevers-bazanella-identifiability-2024}).

\begin{definition}
				A node $j$ is called a \textbf{dource} if it has at least one out-neighbor to which all  its in-neighbors have a directed edge.				\label{def:dource}
				A node $j$ is called a \textbf{dink} if it has at least one in-neighbor that has a directed edge to all  its
  out-neighbors.		
\end{definition}

{\ni \bf Assumptions on the network matrix $G(z)$}\\
Throughout the paper, we shall make the following assumptions on the network matrix:
\begin{itemize}
	\item the diagonal elements are zero and all other elements are  proper;
	\item $(I - G(z))^{-1}$ is proper and all its elements are stable.
\end{itemize}

One can represent the dynamic network in (\ref{eq:dynet1})-(\ref{eq:dynet2}) as an input-output model as follows
\begin{align}
	y(t) = M(z) r(t),\; \text{with}\; M(z) \triangleq C T(z) B.
	\label{eq:IOdynet}
\end{align}
where $T(z) \triangleq (I - G(z))^{-1}$.
Observe that the matrix $T(z)$ is generically nonsingular by construction. 

In analyzing the generic  identifiability of the network matrix, it is assumed that the input-output model $M(z)$ is known; the identification of $M(z)$ from  input-output (IO) data $\lbrace y(t), r(t) \rbrace$ is a standard  identification problem, provided the input signal $r(t)$ is sufficiently rich. The question of generic identifiability of the network is then whether the network matrix $G(z)$ can be fully recovered from the  transfer matrix $M(z)$. 
It is defined as follows.

%
\begin{definition}
(\cite{hendrickx-gevers-bazanella-identifiability-2019}) The network matrix $G(z)$ is generically identifiable from excitation signals applied to $\mathcal{B}$ and measurements made at $\mathcal{C}$ if, for any rational transfer matrix parametrization $G(P, z)$ consistent with the directed digraph associated with $G(z)$, there holds
	\[
					C [I - G(P, z)]^{-1} B = C[I - \tilde{G}(z)]^{-1} B \implies G(P, z) = \tilde{G}(z),
	\]
	for all parameters $P$ except possibly those lying on a zero measure set in $\mathbb{R}^{N}$, where $\tilde{G}(z)$ is any network matrix consistent with the digraph.
\end{definition}
%
The following Proposition provides necessary conditions for generic identifiability of any network; 
it combines Theorem III.1, Corollary III.1 from \cite{bazanella-gevers-hendrickx-network-2019} with Theorem III.1 from \cite{mapurunga-gevers-bazanella-identifiability-2024}. 

\begin{proposition}\label{corol1}
The network matrix $G(z)$ is generically identifiable only if $\mathcal{B}, \mathcal{C} \neq \emptyset$; $\mathcal{F}, \mathcal{D}_F \subset \mathcal{B}$; $\mathcal{S}, \mathcal{D}_S \subset \mathcal{C}$  and $\mathcal{B} \cup \mathcal{C} = \mathcal{V}$.\label{cor:fromBandC}
\end{proposition}
This paper deals with networks in which not all nodes are excited and not all nodes are measured. 
Finding conditions that guarantee generic  identifiability for such  networks  is equivalent to constructing an Excitation and Measurement Pattern (EMP)  that guarantees identifiability.
The concept of EMP and of valid EMP, which led  to the concept of minimal EMP, was introduced in \cite{mapurunga-optimal-2021}. They are defined in the following. 
\begin{definition}
    A pair of selection matrices $B$ and $C$, with its corresponding pair of node sets $\mathcal{B}$ and $\mathcal{C}$,  defines  an \textbf{excitation and measurement pattern (EMP)}.
  An EMP is called \textbf{valid} for the network (\ref{eq:dynet1})-(\ref{eq:dynet2}) if this network is generically identifiable with this EMP.
  Let $\nu = |\mathcal{B}| + |\mathcal{C}|$ \footnote{$|\cdot|$ - Denotes the cardinality of a set.} be the cardinality of an EMP.
  A given EMP is called  \textbf{minimal} for this network if it is valid and there is no other valid EMP with smaller cardinality.
  \label{def:EMP}
\end{definition}
Proposition~\ref{corol1} shows that for an EMP to be valid, it must contain at least one excitation and one measurement, all sources and dources must be excited and all sinks and dinks must be measured, and every other node must be either excited or measured. 
From now on, we drop the arguments $z$ and $t$ used in (\ref{eq:dynet1})-(\ref{eq:dynet2}) whenever there is no risk of confusion. 

\section{Identifiabillity results for specific network structures} \label{specific}

In this section, we present generic identifiability results for some specific network structures, for the situation of partial excitation and measurement. Stated otherwise, we present valid EMPs for these network structures. We consider three types of network structures that are defined by their specific topologies, namely trees, loops and parallel paths networks. We first briefly recall recent existing identifiability results for trees and loops. We then present novel identifiability results for parallel paths networks. First, we  define these three specific network structures.

\subsection{Trees}\label{sec:trees}

\begin{definition}
A directed tree is a weakly connected graph which has no loops even if one were 
 to change the edges directions.
\end{definition}

For the identifiability of directed trees, the following necessary and sufficient conditions were derived in \cite{bazanella-gevers-hendrickx-network-2019}.

 \begin{theorem}\label{treeal}
 A directed tree is generically  identifiable if and only if  
 ${\cal F}\subseteq {\cal B}$,
 ${\cal S}\subseteq {\cal C}$,
 ${\cal B}\cup {\cal C} = {\cal V}$.
 \end{theorem}

%

\subsection{Loops}\label{sec:loops}

\begin{definition}
A loop is a network consisting of a path that starts and ends at the same node. If the loop is part of a larger network, it is called an \it{isolated loop} if no other loop in the graph contains any of the nodes of the loop of interest.
\end{definition}

In \cite{mapurunga-gevers-bazanella-necessary-2022} the following necessary and sufficient conditions were derived for the generic identifiability of isolated loops that have at least three nodes.
 \begin{theorem}\label{NSCcondCor}
All transfer functions in an isolated loop are generically
identifiable if and only if ${\cal B} \cup {\cal C} = {\cal V}$ and, in addition:
(i) either ${\cal{B} \cap \cal{C} }\neq \emptyset$,
 or
(ii) the excited nodes (and hence also the measured nodes) are not all consecutive along the loop. 
\end{theorem}
Alternative versions of necessary and sufficient conditions for the identifiability of isolated loops can be found in  \cite{mapurunga-gevers-bazanella-necessary-2022}, but the formulation given in Theorem~\ref{NSCcondCor} is by far the simplest; the verification can be done by visual inspection of the loop. 

\subsection{Parallel Paths Networks}\label{sec:PPN}

\begin{figure}[h!]
			\centering
			\includegraphics[width=0.8\columnwidth]{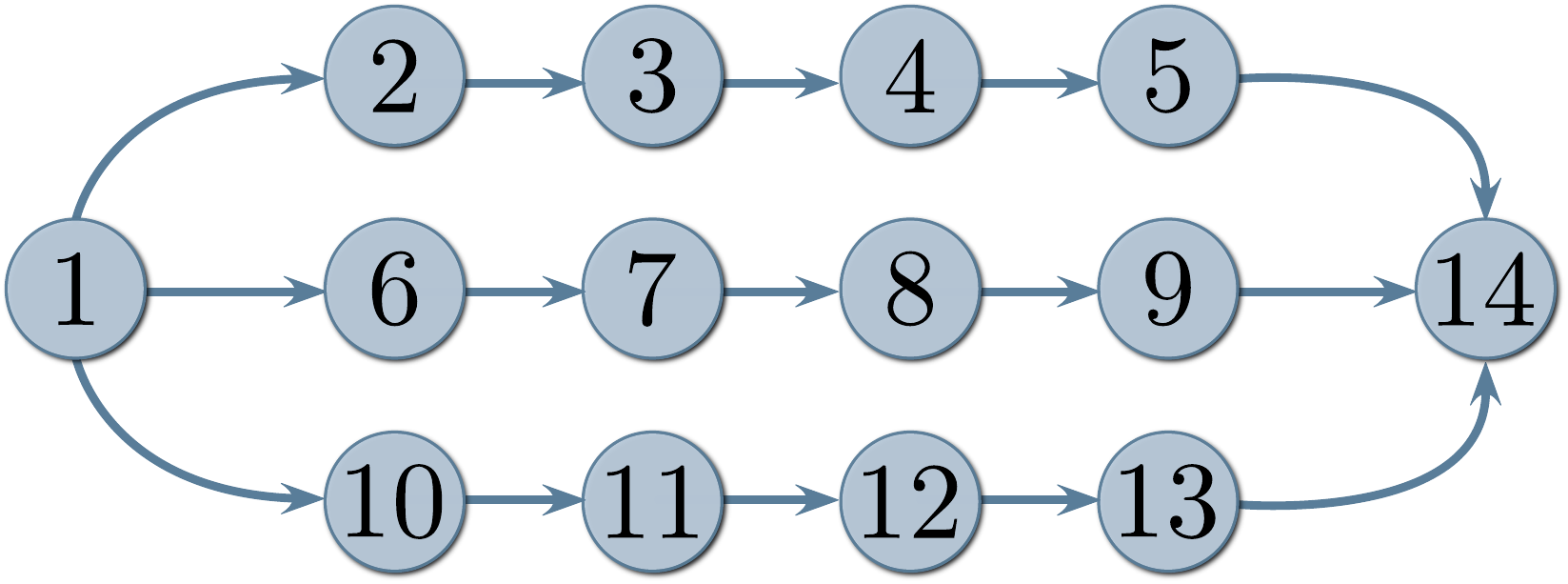}
			\caption{An example of a parallel paths network. }
			\label{fig:parallel}
\end{figure}

We now consider a class of networks with a single source, a single sink, and $n_p$ paths with an arbitrary number of nodes between the source and the sink.
Figure \ref{fig:parallel} depicts an example of such network. 

\begin{definition}
  A \textbf{parallel paths network (PPN)}  is a network composed of a single source and a single sink. 
	There are $n_p \ge 2$ paths from the source to the sink,
	which are the only nodes common to each path.
	At most one path may have no internal nodes; every other path has at least three nodes.
	\label{def:parallel}
\end{definition}
This definition extends the concept of parallel network presented in \cite{mapurunga-identifiability-2021}, where each path contained only a single node. 
The following theorem gives necessary and sufficient conditions for the generic identifiability of a PPN.
\begin{theorem}
				 Consider a parallel paths network from Definition \ref{def:parallel} with 
				one source $\mathcal{F} = \{1\} $ 
				connected through $n_{p} \ge 2$ paths $\left(\mathcal{P}_1, \mathcal{P}_2, \dots, \mathcal{P}_{n_p} \right)$  
				to one sink $\mathcal{S} = \{n\} $.				
								Let $\mathcal{V}_{\mathcal{P}_j}$ be the set of nodes of  path $\mathcal{P}_j$. 
												Assume that the nodes have been labeled sequentially,
				such that for every path $\mathcal{P}_{j}$ there exists a path from $k'_j$ to $k''_j$ only if $k'_j < k''_j$. 
				A parallel paths network is generically identifiable if and only if 
				$\mathcal{F} \subset \mathcal{B}$, 
				$\mathcal{S} \subset \mathcal{C}$, 
				$\mathcal{B} \cup \mathcal{C} = \mathcal{V}$, and in addition
				there are $n_p - 1$ paths $\mathcal{P}_j$ for which there exist at least 
				$k_j', k_j'' \in \mathcal{V}_{\mathcal{P}_j} \setminus \{1, n\} $, such that $k_j' \le k_j''$, $k_j' \in \mathcal{B}$,  $k_j'' \in \mathcal{C}$.     
				\label{theo:generalparallel}
\end{theorem}
\begin{proof}
				\textit{Necessity}: 
				The first three conditions are known to be necessary for any network. 
				It remains to justify the last condition.
				Assume that two paths do not obey the last condition. 
				Without loss of generality, we denote them $\mathcal{P}_1$, $\mathcal{P}_2$. 
				This implies that, for each of these two paths, there are three possibilities:
				1) all \emph{internal} nodes of a particular path are excited, 
				2) all \emph{internal} nodes of a path are  measured, or
				3) the first $l$ \emph{internal} nodes of a path are measured and the remaining are excited. 
				Assume, without loss of generality, that each path $\mathcal{P}_k$ for $k = 1, 2, \dots, n_j$, has $p_{k} + 2$ nodes. 
				Hence, the two paths $\mathcal{P}_1$ and $\mathcal{P}_2$ have a total of $p_1 + p_2 + 2$ unknown edges.
				Now, from Lemma III.1, 
				we know that if there is no path from node $i$ to node $j$ then $T_{ji} = 0$. 
				Thus, for the two paths $\mathcal{P}_1$ and $\mathcal{P}_2$ there are $p_1 + p_2 + 1$ useful equations, 
				because there are $p_1 + p_2$ elements $T_{ji} \neq 0$ and in addition,
				we have $T_{n 1} \neq 0$, which is common to the paths.
				Therefore, there are more unknowns than available useful equations. 

				\textit{Sufficiency}: Let us consider that path $\mathcal{P}_{m}$ is the only path that does not obey the last condition stated in the Theorem; hence  all other paths do obey this condition.
				We now show that 
				for each path $\mathcal{P}_{l} \neq \mathcal{P}_{m}$ all edges can be generically identified. 
				Let $\mathcal{V}_{\mathcal{P}_l} = \{1, k^l_1, k^l_2, \dots, k^l_{p_k}, n\}$, and assume that node $k^l_i$ is excited and node $k^l_j$ is measured, with $k^l_j \ge k^l_i$. 
				The transfer functions corresponding with the edges of $\mathcal{P}_l$ can be identified as follows:
				\begin{align}
								\text{if}\; k^l_1 \in \mathcal{B},~ G_{k^l_1, 1} &= T_{k^l_j, 1} / T_{k^l_j, k^l_1}; \\
								\text{if}\; k^l_1 \in \mathcal{C},~ G_{k^l_1, 1} &= T_{k^l_1, 1}
				.\end{align}
				Consider now $k^l_2$. 
				If $k^l_2 \in \mathcal{C}$ then we can recover $G_{k^l_2, k^l_1} = T_{k^l_2, 1} / G_{k^l_1, 1}$. 
				Conversely, if $k^l_2 \in \mathcal{B}$, then we can recover $G_{k^l_2, k^l_1} = T_{k^l_j, 1}  / (T_{k^l_j, k^l_2} G_{k^l_1, 1})$.  
				We can apply the same reasoning for all nodes $k = k^l_1, k^l_2, \dots, k^l_{j}$ (remember that node $k^l_j$ is measured) and thus recover all transfer functions up to $G_{k_j^l, k_{j-1}^l}$. 
				In a similar fashion we can recover the remaining transfer functions as:
				\begin{align}
								\text{if}\; &k^l_j + 1 \in \mathcal{B},\; G_{k^l_j+1, k^l_j} = T_{n, k^l_i} / (T_{k^l_j, k^l_i} T_{n, k^l_j + 1}); \\
								\text{if}\; &k^l_j + 1 \in \mathcal{C},\; G_{k^l_j+1, k^l_j} =  T_{k^l_j+1, k^l_i} / T_{k^l_j, k^l_i} 
				.\end{align}
				Applying this reasoning to the remaining nodes $k = k^l_j+1, k^l_j+2, \dots, n$ in $\mathcal{P}_l$ one can recover all transfer functions in $\mathcal{P}_l$,
				for all paths other than $\mathcal{P}_m$.

				Now, consider the remaining path $\mathcal{P}_m$. 
				It has $p_m + 2$ nodes corresponding to $\mathcal{V}_{\mathcal{P}_m} = \{1, k^m_1, k^m_2, \dots, k^m_{p_m}, n\}$ and $p_m + 1$ unknown transfer functions. Since we assume that $P_m$ does not obey the last condition of the Theorem, its EMP obeys of the three possible scenarios previously stated. 				
				Suppose that $k_1^m$ is excited.
				It follows what we are in scenario 1, and hence all other nodes of  $\mathcal{P}_m$ need to be excited.
				In this situation, we can recover all transfer functions of $\mathcal{P}_m$, except for $G_{k_1^m, 1}$.
				Suppose now that $k^m_{p_m}$ is measured.
				This means that we are in scenario 2, 
				and hence all other nodes of $\mathcal{P}_m$ are measured.
				In a similar way, we can recover all transfer functions of $\mathcal{P}_m$, except for $G_{n, k_{p_m}^m}$.
				The last scenario is where the first $\{k_1^m, k_2^m, \dots, k_j^m\} $ are measured and the remaining $\{k_{j+1}^m, \dots, k_{p_m}^m\}$ are excited.
				For this case, we can recover all modules of $\mathcal{P}_m$, except for $G_{k_{j+1}^m, k_j^m}$.
				We conclude that, in all cases, all transfer functions in path $\mathcal{P}_m$ can be identified except one.
				Since  all other transfer functions in the network are known, this remaining unknown transfer function can be successfully recovered from $T_{n 1}$, which has not been used for the computation of the other transfer functions.
\end{proof}
What Theorem~\ref{theo:generalparallel} says is that, for a PPN to be generically identifiable, all paths except one must have an excitation that precedes a measurement, in addition to the universal necessary condition that the source must be excited, the sink must be measured, and all other nodes must be either excited or measured. 
This also implies that, for paths that would contain only one internal node, 
these nodes must be both excited and measured.
Notice that this result allows one to identify PPNs with minimal cardinality.

\section{Isolated versus embedded}\label{isolemb}

The necessary and sufficient conditions for generic identifiability for trees, loops and PPNs, respectively, allow one to construct valid and even minimal EMPs for these specific structures. However, these necessary and sufficient conditions, and the EMPs that are based on them, are valid for these structures taken in isolation. 
When these same structures are part of a more  complex network, these EMPs may no longer be valid. To clarify this point, we will now introduce the concept of subdigraph, and specify exactly what we mean by an isolated subdigraph and an embedded subdigraph. We will then illustrate why an EMP that is valid for a subdigraph taken in isolation may no longer be valid when that subdigraph is embedded in a larger network.

Let $\mathcal{V}_A$ be a subset of nodes of $\mathcal{V}$, and 
let $\mathcal{G}_A(\mathcal{V}_A, \mathcal{E}_A)$ be the subdigraph
of $\mathcal{G}$ defined by the subset of nodes $\mathcal{V}_A$ and 
all
the edges $\mathcal{E}_A$ that link them, 
and let $G_A$ be the associated subnetwork matrix.  
We extend Definition \ref{def:EMP} to	deal with subdigraphs. 

\begin{definition}
Consider a digraph $\mathcal{G}(\mathcal{V}, \mathcal{E})$ and a subdigraph $\mathcal{G}_A(\mathcal{V}_A, \mathcal{E}_A)$ of $\mathcal{G}(\mathcal{V}, \mathcal{E})$. Consider an EMP that is valid for the subdigraph $\mathcal{G}_A(\mathcal{V}_A, \mathcal{E}_A)$ when all vertices and edges
from $\mathcal{G}(\mathcal{V}, \mathcal{E})$ that do not belong, respectively,
to $\mathcal{V}_A$ and  $\mathcal{E}_A$ have been removed. This EMP is said to be valid for the isolated subdigraph 
$\mathcal{G}_A(\mathcal{V}_A, \mathcal{E}_A)$. If the same EMP is valid when $\mathcal{G}_A(\mathcal{V}_A, \mathcal{E}_A)$ is embedded into $\mathcal{G}(\mathcal{V}, \mathcal{E})$, then this EMP is said to be valid for the embedded $\mathcal{G}_A(\mathcal{V}_A, \mathcal{E}_A)$.
\label{def:EMPsubemb}
\end{definition}

Finding a valid, possibly minimal, EMP for the isolated graph is likely to be a much simpler problem
				than finding an EMP that is valid for the embedded graph. We illustrate this with the following example.\\

\begin{figure}[h!]
				\centering
				\includegraphics[width=0.4\columnwidth, keepaspectratio]{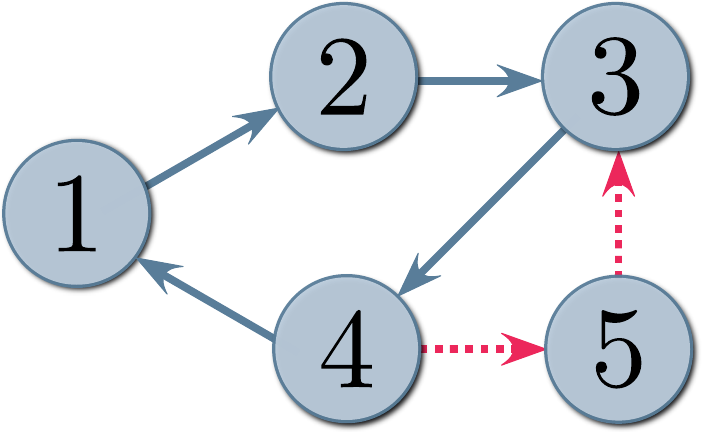}
				\caption{A network formed by two loops: the one on the left formed by the solid blue edges, and the other on the right formed by the dotted red edges and the edge $3\to4$.}
				\label{fig:Ex1-loops}
\end{figure}

\ni{\bf Example 1}\\
Consider the 5-node network depicted in Figure~\ref{fig:Ex1-loops}, with network matrix given by
\begin{align}
			G = 
			\left[\begin{matrix}
							0 & 0 & 0 & G_{14} & 0\\
							G_{21} & 0 & 0 & 0 & 0\\
							0 & G_{32} & 0 & 0 & G_{35}\\
							0 & 0 & G_{43} & 0 & 0\\
							0 & 0 & 0 & G_{54} & 0
			\end{matrix}\right]
			\label{eq:Ex-1-G}
.\end{align}


Suppose first that the dotted red edges are not present in this network ($G_{54} = G_{35} = 0$), 
and that our objective is to identify the loop formed by nodes 1, 2, 3 and 4. 
According to Theorem \ref{NSCcondCor}, there are two minimal EMPs for this isolated loop, namely:
$\text{EMP}_1: \mathcal{B}_1 = \{1, 3\}, \mathcal{C}_1 = \{2, 4\} $ and 
$\text{EMP}_2: \mathcal{B}_2 = \{2, 4\}, \mathcal{C}_2 = \{1, 3\} $.
However, when this isolated loop is connected to the other loop - and hence embedded in the $5$-node network -  as seen in Figure \ref{fig:Ex1-loops}, 
these EMPs are no longer valid for this embedded loop. 
This example shows that a valid EMP for an isolated digraph may no longer be valid when this digraph is embedded into a more complex network. 
Therefore, using  the theorems of Section \ref{specific} to produce valid EMPs for subdigraphs with specific structures may no  longer yield  valid EMPs when these subdigraphs are embedded in a larger digraph.
In the next section we present our main result which exhibits conditions under which an EMP that is valid for an isolated digraph remains valid when it is embedded into a larger network.

\section{Main result}\label{mainresult}

\begin{theorem}
				Consider a  digraph $\mathcal{G}(\mathcal{V}, \mathcal{E})$ and its associated
				network matrix $G$.
				Let $\mathcal{V}_A$ be a subset of connected nodes of $\mathcal{V}$,
				and ${\mathcal{V}_B} \triangleq \mathcal{V}\setminus \mathcal{V}_A$, 
				Let $\mathcal{G}_A(\mathcal{V}_A, \mathcal{E}_A)$ be the subdigraph of $\mathcal{G}$ defined by the subset of nodes $\mathcal{V}_A$ and 
				all edges $\mathcal{E}_A$ that link them, 
				and let $G_A$ be the associated subnetwork matrix.
				Assume that an EMP that is valid for the isolated subdigraph $\mathcal{G}_A$ has been obtained.
				This same EMP is also valid for the embedded subdigraph $\mathcal{G}_A$
				if  at least one of the following conditions holds.
				\begin{enumerate}
								\item There is no path starting in $\mathcal{V}_A$, passing through $\mathcal{V}_B$, and returning to $\mathcal{V}_A$. 
								\label{theo:item:zero-path}
								\item All paths that leave
								$\mathcal{V}_A$ and return to $\mathcal{V}_A$ are known.
								\label{theo:item:known-path}
							\end{enumerate}
				\label{theo:no-path-decomp}
\end{theorem}
\begin{proof}
				Let us start by partitioning the network matrix $G$ and the corresponding input-output matrix $T$ according to the sets $\mathcal{V}_A$ and $\mathcal{V}_B$.
				\begin{align*}
								G = 
								\begin{bmatrix} 
												G_{A}   & G_{A B} \\
												G_{B A} & G_{B}
								\end{bmatrix} 
								,
								T = 
								\begin{bmatrix} 
												T_{A A} & T_{A B} \\
												T_{B A} & T_{B B}
								\end{bmatrix} 
				.\end{align*}
				The choice of an EMP for  $\mathcal{G}_A$   corresponds 
to the choice of a submatrix of $T_{A A}$, namely the submatrix defined by the corresponding excited and measured 
nodes of $\mathcal{V}_A$. This EMP is valid if $G_{A}$ can be uniquely reconstructed from this submatrix of $T_{A A}$.  To examine whether an EMP that is valid for the identification of $G_{A}$, taken in isolation, remains  valid when the nodes $\mathcal{V}_A$ are connected to the nodes $\mathcal{V}_B$, we use the following identity that relates  $T_{A A}$  to the other submatrices:
				\begin{align}
								T_{A A} &= (I_A - G_A - G_{A B} (I_{B} - G_{B})^{-1} G_{B A})^{-1} \nonumber \\ 
										    &= (I_A - G_A - G_{A B} T_{B } G_{B A})^{-1}
							  \label{eq:TAA}
				,\end{align}
				where $T_{B} \triangleq (I - G_{B})^{-1}$. 		
								All elements of $G_{A B} T_{B} G_{B A}$ can be written as:
				\begin{align*}
								\sum_{k_1, k_2 \in \mathcal{V}_A, j, i \in \mathcal{V}_B} G_{k_1 j} T_{j i} G_{i k_2} 
				.\end{align*}
				Now, if there is no path from $\mathcal{V}_A$ passing through $\mathcal{V}_B$ and returning to $\mathcal{V}_A$, then one of the following holds:
				1) $G_{k_1 j} = 0$; 
				2) $T_{j i} = 0$;
				3) $ G_{i k_2} = 0$,
				for all $k_1, k_2 \in \mathcal{V}_A$ and $j, i \in \mathcal{V}_B$.
				Hence, all elements of $G_{A B} T_{B} G_{B A}$ are zero, which implies that 
				\begin{align}
					T_{A A} = (I_A - G_A)^{-1} \triangleq T_{A}
				\label{eq:TAAisolated}
				\end{align}
				This is the expression that relates  $G_A $ to $T_A$ in the subnetwork $\mathcal{G}_A$  taken in isolation. We have shown that the same relationship holds when that subnetwork is embedded in the complete network.
				Hence, the network matrix $G_A$ can be recovered if an EMP is applied to $T_{AA}$ that is  valid for the subnetwork  $\mathcal{G}_A$ taken in isolation, which  proves item \ref{theo:item:zero-path}.\\
				In order to prove item \ref{theo:item:known-path}, we note that the expression (\ref{eq:TAAisolated}) relates $T_{AA}$ to $G_A$ for the subnetwork $\mathcal{G}_A$ taken in isolation. If an EMP is valid for this subnetwork taken in isolation, it means that, with the  submatrix of $T_{AA}$ corresponding to this EMP, $G_A$ can be recovered from  $G_A = I_A - T_{A A}^{-1}$. When this subnetwork is embedded in the full network, this  relation becomes (using a rewrite of (\ref{eq:TAA})):
				\begin{align}
								G_A = I_A - T_{A A}^{-1} - G_{AB} T_B G_{BA}
								\label{eq:GAid}
				.\end{align}
				If all paths leaving $\mathcal{V}_A$ and returning to $\mathcal{V}_A$ are known, then $G_{A B} T_{B} G_{B A}$ is also known. 
				Hence, $G_A$ can also be identified from (\ref{eq:GAid}), and this completes the proof. 
\end{proof}

Theorem~\ref{theo:no-path-decomp} is a powerful result for the situation where only part of the digraph
is of interest, and where the task is to design a valid EMP for the corresponding subdigraph. But it also allows one to synthesize EMPs for the full network if a convenient decomposition of the digraph can be obtained. We illustrate this with the following Example.\\

\ni{\bf Example 2}

\begin{figure}[h!]
			\centering
			\includegraphics[width=0.8\columnwidth]{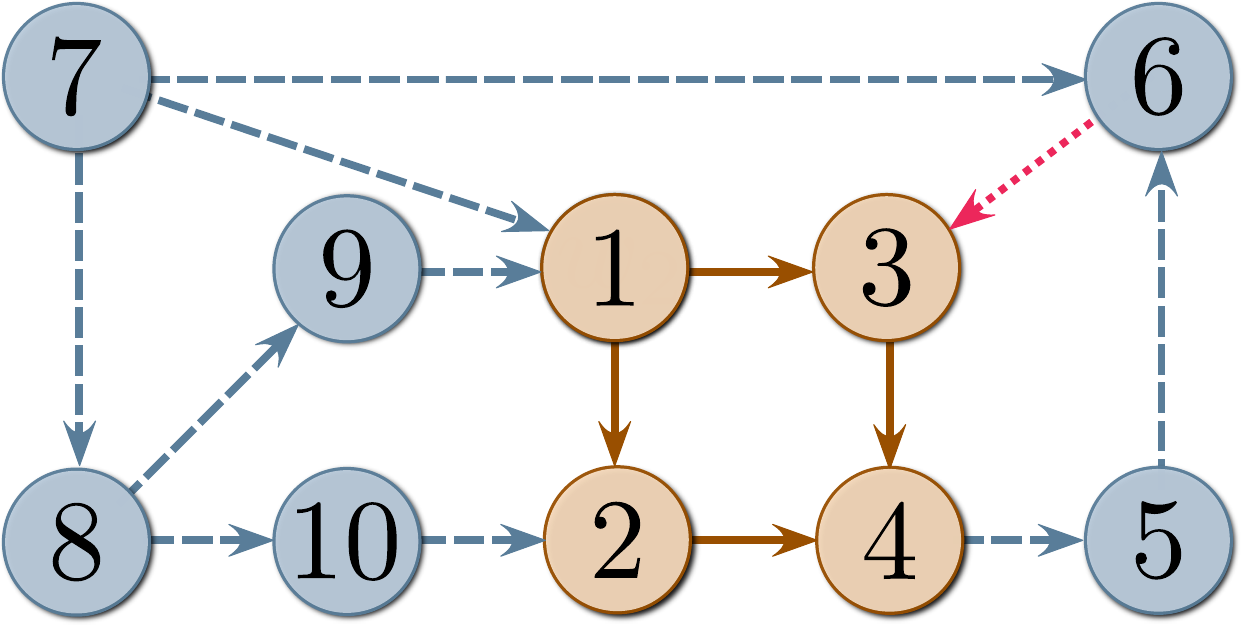}
			\caption{
											A network with multiple strucures. 
											The nodes $1, 2, 3, 4$ form a PPN.
											The  nodes $3, 4, 5, 6$ form a loop. 
											The dashed blue edges linking nodes $1, 2, 6, 7, 8, 9, 10$ form a tree.}
			\label{fig:10}
\end{figure}

Consider the network in Figure \ref{fig:10}, with ten nodes and twelve edges. Assume first that the edge from $6$ to $3$ does not exist, 
i.e.  disregard the dotted red edge for now.
Suppose first that one only wants to identify the subdigraph, denoted $\mathcal{G}_A(\mathcal{V}_A, \mathcal{E}_A)$, that connects nodes $1$ to $4$, i.e. $\mathcal{V}_A= \{1,2,3,4\}$. 
It is a PPN.
According to Theorem \ref{theo:generalparallel},
the following EMP, denoted $\text{EMP}_1$,
is valid for this subdigraph taken in isolation:  
${\cal B}_1 = \{1, 3\}$, ${\cal C}_1 = \{2, 3, 4\}$; it is actually minimal. 
Now, consider that same subdigraph $\mathcal{G}_A(\mathcal{V}_A, \mathcal{E}_A)$ embedded in the whole digraph formed by all ten  nodes shown in Figure~\ref{fig:10}, still without the edge $6\to 3$. We observe that this subdigraph obeys  condition 1 of Theorem~\ref{theo:no-path-decomp}, and hence \emp{1} remains valid for the identification of the embedded subdigraph. 

Now consider that the dotted edge $6\to 3$ (i.e. $G_{36}$) 
is added to the previous graph. 
Then condition 1 of Theorem~\ref{theo:no-path-decomp} is no longer satisfied, since there is an unknown path leaving $\mathcal{V}_A$ and returning to itself: 
$( 4\to 5\to 6 \to 3)$ $\{G_{54}, G_{65}, G_{36}\}$. 
If these edges are known, then condition 2 of Theorem~\ref{theo:no-path-decomp} will be valid for the identification of the embedded $\mathcal{G}_A$.
 So, one possibility to  produce a valid EMP for the identification of all edges
 in $\mathcal{G}_A$, when it is embedded in  the complete network $\mathcal{G}$, is to identify also the edges in this path.

This path, together with $3\to 4$  $(G_{43})$, forms a  loop: $\{3, 4, 5, 6, 3\}$. 
Consider first this loop in isolation. 
It then follows from Theorem~\ref{NSCcondCor} that the following EMP, denoted $\text{EMP}_2$,  allows the identification of all edges in this isolated loop: 
${\cal B}_2 = \{3, 5\}$, ${\cal C}_2 = \{4, 6\}$. 
Now, consider this loop as embedded in the complete network $\mathcal{G}$. 
We observe that there is no path in $\mathcal{G}$ that leaves this loop and re-enters it. 
Hence, the conditions of item 1) of Theorem~\ref{theo:no-path-decomp} apply, 
which means that this $\text{EMP}_2$  is also valid for this loop $\{3,4,5,6,3\}$ when it is embedded in the complete network $\mathcal{G}$. 

Combining $\text{EMP}_1$ and $\text{EMP}_2$,  
we have $\text{EMP}_3$, defined as ${\cal B}_3 = \{1, 3, 5\}$, ${\cal C}_3 = \{2, 3, 4, 6\}$, 
which allows to identify all edges in $\mathcal{G}_A$ 
when it is embedded in the whole network shown in Figure~\ref{fig:10}, including node $6$. 
This illustrates how Theorem~\ref{theo:no-path-decomp} can be used to construct a valid EMP for a subdigraph that is embedded in a more complete digraph. 
As a byproduct, $\text{EMP}_3$  also allows the identification of the transfer functions $G_{54}, G_{65}, G_{36}$.

Finally, consider that one wants to identify the whole network, that is,
all the edges in $\mathcal{G}$. Let $\mathcal{G}_C$ be the graph formed
by the nodes $\mathcal{V}_C = \{1, 2, 6, 7, 8, 9, 10\} $ and
by the edges that are not identified with $\text{EMP}_3$, that is,
$\mathcal{E}_C = \{7\to 6, 7\to 1, 7\to 8, 8\to 9, 9 \to 1, 8 \to 10, 10 \to 2\}$.
Actually,  $\mathcal{G}_C$ is a tree, so a  valid EMP for it in isolation is
${\cal B}_4 = \{7, 8, 9,10\}$, ${\cal C}_4 = \{1, 2, 6\}$, denoted $\text{EMP}_4$.
If all edges in $\mathcal{G}_A$, as well as  the edges $4\to 5, 5\to 6, 6\to 3$ 
have been identified using \emp{3}, condition 2) of Theorem~\ref{theo:no-path-decomp} is satisfied for $\mathcal{G}_C$.
 Therefore, the combination of $\text{EMP}_3$ and $\text{EMP}_4$ allows the identification of the whole network. 
 Define it as $\text{EMP}_5$: 
 ${\cal B}_5 = \{1, 3, 5, 7, 8, 9,10\}, 
 {\cal C}_5 = \{1, 2, 3, 4, 6 \}$. 
 Its  cardinality is $12$, much smaller than the maximum $19$ - the graph has only one source and 
no sinks - and only slightly more than the minimum $10$.
Notice that different combinations of valid EMPs for each subnetwork can be used to produce a range of EMPs that are valid for the identification of the whole network.

Example 2 has allowed us to illustrate the main result of Theorem~\ref{theo:no-path-decomp}. 
It has shown how a valid, possibly minimal, EMP constructed for a subdigraph taken in isolation can be preserved or enhanced when that subdigraph is embedded in a larger digraph. 
At the same time, Example 2 has illustrated that, when a network is  decomposed into a set of simple subdigraphs for which valid EMPs are easy to construct, 
the combination of these valid EMPs can lead to an EMP that is valid for the whole digraph,
via the use of  Theorem~\ref{theo:no-path-decomp}. 
Our ongoing work is to develop an algorithm for the decomposition of a general network into subdigraphs to which this scenario can be applied.

\section{Conclusion}
We have presented new results for the generic identifiability of Parallel Paths Networks. But our main contribution has been to show that an EMP that is valid for a subnetwork treated in isolation is typically no longer valid when that subnetwork is embedded in a larger network, and to present sets of sufficient conditions under which it remains valid. This result is of importance for the practical situation when only part of a network needs to be identified. With a proper decomposition of the global network, it may also lead to an efficient method for the identification of this global network.

\bibliographystyle{IEEETran}
\bibliography{decbib}

\end{document}